\documentclass[useAMS,usenatbib,usegraphicx]{mn2e}
\usepackage{times}

\newcommand{\aj}{AJ}                   
\newcommand{\araa}{ARA\&A}             
\newcommand{\apj}{ApJ}                 
\newcommand{\apjl}{ApJL}               
\newcommand{\apjs}{ApJS}               
\newcommand{\aap}{A\&A}                
\newcommand{\aapr}{A\&AR}              
\newcommand{\mnras}{MNRAS}             
\newcommand{\pasa}{PASA}               
\newcommand{\pasp}{PASP}               

\title[The gas-to-dust mass ratio of Cen~A as seen by \emph{Herschel}]{The gas-to-dust mass ratio of Centaurus~A as seen by \emph{Herschel}\thanks{\emph{Herschel} is an ESA space observatory with science instruments provided by European-led Principal Investigator consortia and with important participation from NASA.}}
\author[T. J. Parkin et al.]{T.~J.~Parkin,$^{1}$\thanks{E-mail: parkintj@mcmaster.ca} C.~D.~Wilson,$^{1}$ K.~Foyle,$^{1}$ M.~Baes,$^{2}$ G.~J.~Bendo,$^{3}$ A.~Boselli,$^{4}$   
\newauthor
M.~Boquien,$^{4}$ A.~Cooray,$^{5}$ D.~Cormier,$^{6}$ J.~I.~Davies,$^{7}$ S.~A.~Eales,$^{7}$
\newauthor
 M.~Galametz,$^{8}$ H.~L.~Gomez,$^{7}$ V.~Lebouteiller,$^{6}$ S.~Madden,$^{6}$ E.~Mentuch,$^{1}$
\newauthor
 M.~J.~Page,$^{9}$ M.~Pohlen,$^{7}$ A.~Remy,$^{6}$ H.~Roussel,$^{10}$ M.~Sauvage,$^{6}$ 
\newauthor
 M.~W.~L.~Smith,$^{7}$ L.~Spinoglio$^{11}$\\
    $^{1}$Dept. of Physics \& Astronomy, McMaster University, Hamilton, Ontario, L8S~4M1, Canada\\
    $^{2}$Sterrenkundig Observatorium, Universiteit Gent, Krijgslaan 281 S9,  
    B-9000 Gent, Belgium \\
    $^{3}$UK ALMA Regional Centre Node, Jodrell Bank Centre for Astrophysics, School of Physics and Astronomy, \\
    University of Manchester, Oxford Road, Manchester M13 9PL, United Kingdom \\
    $^{4}$Laboratoire d'Astrophysique de Marseille - LAM, Universit\'e d'Aix-Marseille \& CNRS, UMR7326, \\ 38 rue F. Joliot-Curie, 13388 Marseille Cedex 13, France  \\
    $^{5}$Department of Physics \& Astronomy, University of California, Irvine,
    CA 92697, USA \\
    $^{6}$CEA, Laboratoire AIM, Irfu/SAp, Orme des Merisiers, F-91191 Gif-sur-Yvette, France \\
    $^{7}$School of Physics and Astronomy, Cardiff University, Queens Buildings The Parade, Cardiff CF24 3AA, UK \\
    $^{8}$Institute of Astronomy, University of Cambridge, Madingley Road, Cambridge, CB3 0HA, UK \\
    $^{9}$Mullard Space Science Laboratory, University College London, Holmbury St Mary, Dorking, Surrey RH5 6NT, UK\\
    $^{10}$Institut d'Astrophysique de Paris, UMR7095 CNRS, Universit\'e Pierre \& Marie Curie, 98 bis Boulevard Arago, \\
    F-75014 Paris, France \\    
    $^{11}$Istituto di Fisica dello Spazio Interplanetario, INAF, Via del Fosso  
    del Cavaliere 100, I-00133 Roma, Italy \\
}

\begin{document}
 
\date{Accepted 0000 January 3. Received 0000 January 2; in original form 0000 January 1}

\pagerange{\pageref{firstpage}--\pageref{lastpage}} \pubyear{2011}

\maketitle

\label{firstpage}

\begin{abstract}
We present photometry of the nearby galaxy NGC~5128 (Centaurus~A) observed with the PACS and SPIRE instruments on board the \emph{Herschel Space Observatory}, at 70, 160, 250, 350 and 500~$\mu$m, as well as new CO~$J=3-2$ observations taken with the HARP-B instrument on the JCMT.  Using a single component modified blackbody, we model the dust spectral energy distribution within the disk of the galaxy using all five \emph{Herschel} wavebands, and find dust temperatures of $\sim 30$~K towards the centre of the disk and a smoothly decreasing trend to $\sim 20$~K with increasing radius.  We find a total dust mass of $(1.59 \pm 0.05) \times 10^{7}$~M$_{\odot}$, and a total gas mass of $(2.7 \pm 0.2)\times 10^{9}$~M$_{\odot}$.  The average gas-to-dust mass ratio is $103 \pm 8$ but we find an interesting increase in this ratio to approximately 275 toward the centre of Cen~A.  We discuss several possible physical processes that may be causing this effect, including dust sputtering, jet entrainment and systematic variables such as the X$_{\mathrm{CO}}$ factor.  Dust sputtering by X-rays originating in the AGN or the removal of dust by the jets are our most favoured explanations.
\end{abstract}


\section{Introduction}\label{sec:intro}
Elliptical galaxies were long thought to be gas and dust poor, but in
the past few decades, several observational surveys 
have shown that a significant fraction of these galaxies contain a rich interstellar
medium (ISM).  Detections of cold dust (temperatures of $\sim 30$~K) in non-peculiar early type galaxies using data from IRAS could only be made in about 12~percent of the sample \citep[e.g.][]{1998ApJ...499..670B}; with improvements in our ability to detect the cold dust component of the ISM, we are finding more evidence for dust within early-types \citep[e.g.][]{2004ApJS..151..237T, 2004A&A...416...41X,2008ApJ...677..249L, 2009AJ....137.3053Y, Smith_2012_in_press}.  Other studies suggest there is a relationship between a detection of cold
dust in these galaxies and the likelihood of a radio source at the core \citep[e.g.][]{1989ApJ...337..209W}.  Cold gas has also been found in elliptical galaxies, both molecular \citep[i.e.][]{1989ApJ...344..747T, 1995A&A...297..643W, 2003ApJ...584..260W, 2011MNRAS.414..940Y}, and atomic \citep[e.g.][]{1985AJ.....90..454K, 1992ApJ...387..484B, 1994A&A...286..389H, 1995A&A...300..675H, 1999ASPC..163...84M, 2006MNRAS.371..157M, 2010ApJ...725..100W}.

Centaurus~A (Cen~A; NGC~5128) is the closest example of a giant elliptical galaxy, located at a distance of $3.8 \pm 0.1$~Mpc \citep{2010PASA...27..457H}. It has a warped disk, likely the result of a past merger event, which contains an abundance of both gas and dust, and has a central active galactic nucleus (AGN) with radio jets extending out to large distances in both directions.  Thus, Cen~A is an ideal laboratory for studying variations in the ISM at high-resolution and to look for effects the AGN might have on the neighbouring disk.  Note that while these interesting traits do make Cen~A a somewhat peculiar elliptical galaxy, some studies show that overall Cen~A is still a normal elliptical on global scales, and that its proximity is what gives us insight to its peculiarity \citep[e.g.][]{2010PASA...27..475H}.

Cen~A has been observed extensively in the past over a wide range of wavelengths.  For a comprehensive review of the numerous physical characteristics of Cen~A, see \citet{1998A&ARv...8..237I} and \citet{2010PASA...27..463M}.  Observations of H~\textsc{i} covering a large area of the sky surrounding Cen~A \citep[e.g][]{1990AJ.....99.1781V,1994ApJ...423L.101S,2008A&A...485L...5M,2010A&A...515A..67S} reveal emission throughout the disk and in shells at large distances from the disk, supporting the merger scenario.  The central core itself shows both blueshifted and redshifted H\textsc{i} absorption \citep{2008A&A...485L...5M,2010A&A...515A..67S}, consistent with a circumnuclear disk.

Due to its low latitude in the southern sky (13$^{\mathrm{h}}$25$^{\mathrm{m}}$27.6$^{\mathrm{s}}$, $-43^{\circ}$01'09''), only a small number of telescopes are capable of observing Cen~A in CO, such as the Swedish-ESO Submillimeter Telescope \citep[SEST; e.g.][]{1987ApJ...322L..73P,1990ApJ...363..451E,1992ApJ...391..121Q,1993A&A...270L..13R}.  Detections in the $J=1-0$ transition by \citet{1990ApJ...363..451E} show that the emission primarily traces the optical disk, but drops off significantly beyond about 90~arcsec on either side of the centre of the disk. In addition, \citet{1990ApJ...363..451E} and \citet{1990ApJ...365..522E} report detections of the $^{12}$CO and $^{13}$CO $J=1-0$ line in absorption against the nuclear non-thermal continuum, and suggest the same molecular clouds are responsible for both the emission and absorption features.

\emph{Spitzer Space Observatory} IRAC observations of Cen~A \citep[][]{2006ApJ...645.1092Q} reveal a prominent parallelogram-shaped ring of dusty material in the central disk at higher resolution than earlier mid-infrared observations of Cen~A using ISOCAM on the \emph{Infrared Space Observatory} \citep[ISO; ][]{1999A&A...341..667M}.  Observations of Cen~A at 450 and 850~$\mu$m with SCUBA and 870~$\mu$m with LABOCA show a warped `S'-shaped structure close to the nucleus spanning approximately 5~arcmin across, similar to the mid-infrared observations \citep{2002ApJ...565..131L,2008A&A...490...77W}.  Less-defined, fainter emission surrounds this structure and extends over much of the optical dust lane.  To explain the `S'-shaped disk, \citet{2006ApJ...645.1092Q} model the warped morphology of the inner infrared/submm disk using a 'tilted-ring' model, consisting of a series of concentric rings.  The IRAS measurements of \citet{1990ApJ...363..451E} reveal an average dust temperature of about 42~K, with the temperature decreasing towards the edge of the disk.  \citet{2002ApJ...565..131L} also find a temperature of $\sim 40$~K in the disk that decreases outward.  Furthermore, \citet{2008A&A...490...77W} show the first detections at submm wavelengths of non-thermal emission from the radio lobes.  Studying the thermal emission within the disk, their two-component dust model shows the dust has temperatures of 30 and 20~K (warm and cold components, respectively), but no decreasing trend at larger radii.  Lastly, a complementary study to this paper by \citet{Auld_2011_submit} analyses the dust properties of the remains of the smaller merger component in Cen~A at large radii from the centre, using data from the \emph{Herschel Space Observatory} \citep{2010A&A...518L...1P}.

In spite of the number of observations of the disk of Cen~A in both the dust and gas components, there are few published values of the gas-to-dust ratio in this galaxy.  \citet{2004A&A...415...95S} report a ratio of $\sim 300$ in a region north of the disk using data taken with ISOPHOT on ISO, and \citet{1998A&ARv...8..237I} suggests an upper limit of 450.  Even for other early-type galaxies, only a small number of surveys report gas-to-dust ratios.  The SINGS survey found ratios similar to that of the Milky Way \citep{2007ApJ...663..866D}, while \citet{2008ApJ...677..249L} found a range of values between $\sim 230$ and 400 for the inner H$_{2}$-dominated regions of seven ellipticals.  In comparison the typical value for the Milky Way is $\sim 160$ \citep[e.g.][]{2004ApJS..152..211Z}.  With \emph{Herschel} we have the capability to study the gas-to-dust ratio for the disk of Cen~A at unprecedented resolution, and we can investigate how this ratio varies throughout the galaxy including regions near the AGN, and try to understand how the AGN affects the structure of the ISM.  \emph{Herschel} also gives us increased sensitivity compared to ground based observatories, and extends coverage to longer wavelengths than previous space missions, which are crucial to measuring the dust content of the ISM.

Here we present a detailed analysis of the gas-to-dust mass ratio of Cen~A through the central disk.  We use photometric observations with both the Photodetector Array Camera and Spectrometer \citep[PACS;][]{2010A&A...518L...2P} and the Spectral and Photometric Imaging Receiver \citep[SPIRE;][]{2010A&A...518L...3G} on \emph{Herschel}, as well as observations with the HARP-B receiver mounted on the James Clerk Maxwell Telescope (JCMT). In Section~\ref{sec:obs} we present the observations and data reduction process.  In Section~\ref{sec:results} we present our \emph{Herschel} and JCMT photometry, our spectral energy distribution (SED) model and maps of the best-fitting dust temperatures, dust and gas masses, and the gas-to-dust ratio across the disk. In Section~\ref{sec:discuss} we discuss the possible mechanisms that might explain the characteristics of our maps, and we conclude in Section~\ref{sec:conclusions}. 

\section{Observations}\label{sec:obs}
\subsection{\emph{Herschel} Observations}\label{subsec:Herschel}
We have obtained PACS photometry at 70 and 160~$\mu$m (OBSIDs 1342188855 and 1342188856), and SPIRE photometry at 250, 350 and 500~$\mu$m (OBSID 1342188663) as part of the Very Nearby Galaxies Survey (VNGS; P.I.: Christine Wilson), which is a \emph{Herschel} Guaranteed Time program.  All of these images cover a 37~arcmin~$\times$~37~arcmin area on the sky centred on Cen~A.  In Table~\ref{table: herschel_char} we summarise the basic characteristics of the original \emph{Herschel} maps, which are shown in Figure~\ref{fig:Herschel_maps}.

\begin{table*}
\begin{minipage}{2\columnwidth}
 \centering
 \caption{A summary of the \emph{Herschel} observation information as well as the pixel noise and calibration uncertainty values.  The measurement uncertainty is pixel-dependent (see Section~\ref{subsec:Noise} for details) and thus values are not presented here.}
 \label{table: herschel_char}
 \begin{tabular}{@{}lrrrrr@{}}
  \hline
  Wavelength & Beam Size & Pixel Size & Colour Correction\footnote{PACS values are divisive and SPIRE values are multiplicative.}  & Pixel Noise\footnote{These uncertainties are for the pixels at their native pixel scale, as listed in this table.}  & Calibration Uncertainty\footnote{We have ignored the fact that the SPIRE calibration errors are correlated between all three bands and our total 7~percent error comprises 5~percent correlated error and 5~percent un-correlated error.} \\
  ($\mu$m)   & (arcsec)  & (arcsec)   &                   & (mJy~pix$^{-1}$)  & (percent)       \\
  \hline
  70         & 5.76      & 2          & 1.043             & 0.009           & 3         \\
  160        & 12.13     & 4          & 1.029             & 0.060           & 5      \\
  250        & 18.2      & 6          & 0.9836            & 0.147           & 7  \\
  350        & 24.5      & 8          & 0.9821            & 0.114           & 7 \\
  500        & 36.0      & 12         & 0.9880            & 0.110           & 7 \\
  \hline
 \end{tabular}
\end{minipage}
\end{table*}

\subsubsection{PACS Observations}\label{subsubsec:PACS}
The PACS 70 and 160~$\mu$m observations were obtained simultaneously in scan-map mode at the `medium' scan speed of 20~arcsec~s$^{-1}$.  The scans were carried out in orthogonal directions to produce a square map with homogenous coverage, with a scan-leg length of 37.0~arcmin.  The observations were then processed using a pipeline that has been modified from the official one released with the \emph{Herschel} Interactive Processing Environment \citep[HIPE, version 4.0;][]{2010ASPC..434..139O}, using calibration file set (FM,v4).  Reduction steps include protecting the data against possible cross-talk effects between bolometers in the array, and deglitching only at the second level.  This involves comparing the signal detected within all of the pixels in the data cube that cover the same location in the sky, and flagging any pixels that are outliers as glitches.  In addition, flat-field corrections were applied to the images, and the signal was converted to units of Jy~pix$^{-1}$.  We also applied correction factors of 1.119 and 1.174 to the 70 and 160~$\mu$m maps, respectively, to update the images with the most recent set of calibration files.  Once these steps were done in HIPE, the final map-making was done in \textsc{Scanamorphos}\footnote{http://www2.iap.fr/users/roussel/herschel/} (version 6.0) \citep{Roussel_2011_submit} using standard settings and pixel sizes of 2.0 and 4.0~arcsec for the 70 and 160~$\mu$m images, respectively.  We applied a colour correction to each map to accommodate the non-monochromatic nature of the galaxy's spectral energy distribution (SED) through each of the PACS filters.  To do so, we first measure the slope of the spectral energy distribution (SED) at 70 and 160~$\mu$m using our one-component blackbody fit (see Section~\ref{subsec:sed} for details), and applied the colour corrections for those slopes.  Then we re-calculate the slope of the SED for the corrected fluxes and check for any change in the slope.  We find slopes of -3 and +1 at 70 and 160~$\mu$m, respectively, and thus we divide these images by correction factors of 1.043 and 1.029 \citep{PACS_cc_2011}.  Lastly, we determine the mean sky value within numerous apertures overlaid onto off-target regions surrounding the galaxy at each wavelength, and then determine the average mean across all apertures.  We take this mean as our background contribution and subtract it from each PACS image.

The final 70 and 160~$\mu$m images have been convolved with appropriate kernels (see \citet{Bendo_2011_submit} for details) to match the 36~arcsec resolution of the 500~$\mu$m image, and then regridded to a 12~arcsec pixel scale, again to match the 500~$\mu$m image.

\subsubsection{SPIRE Observations}\label{subsubsec:SPIRE}
\begin{figure*}
\includegraphics[width=7.75cm]{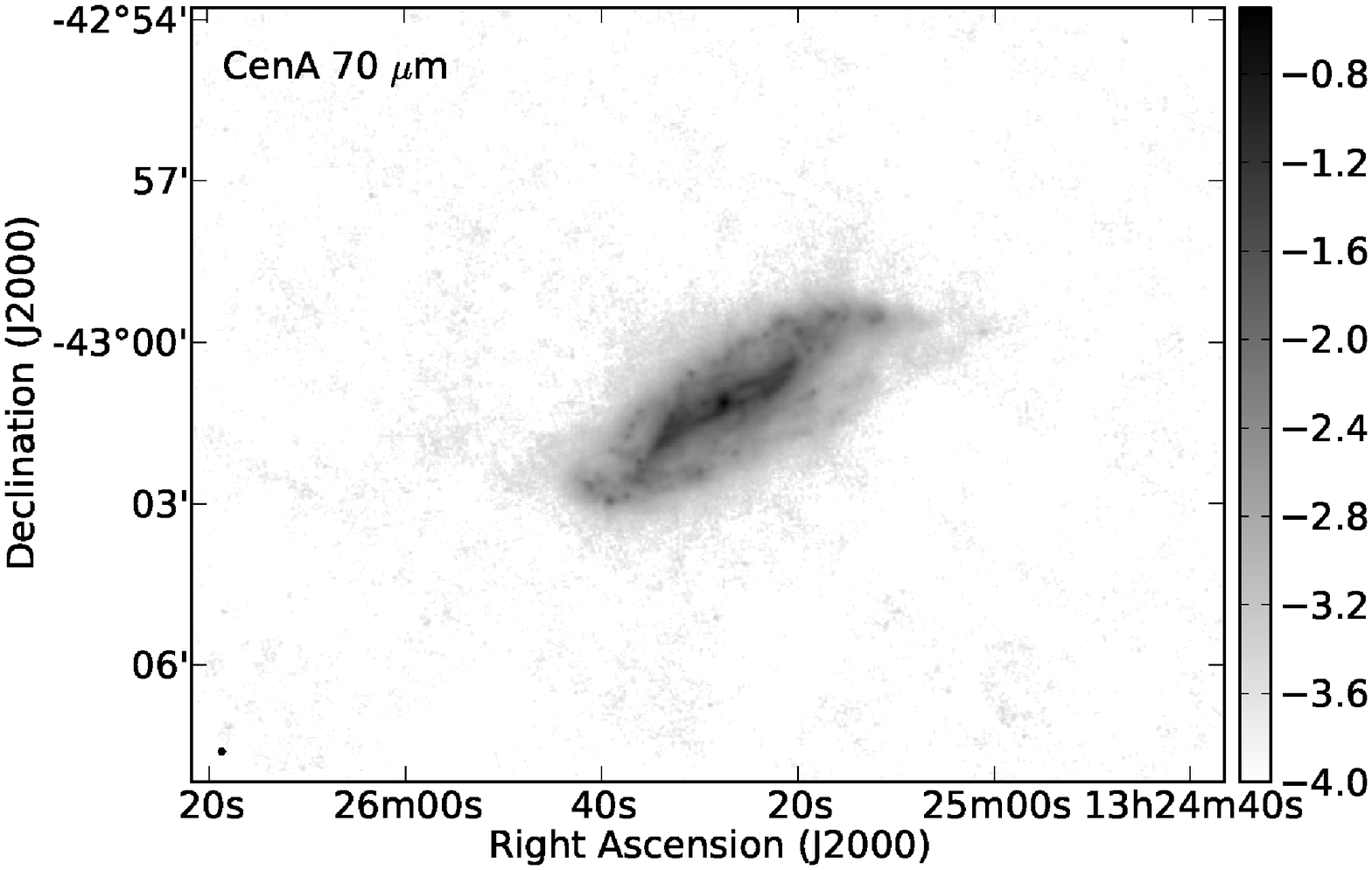}
\includegraphics[width=7.75cm]{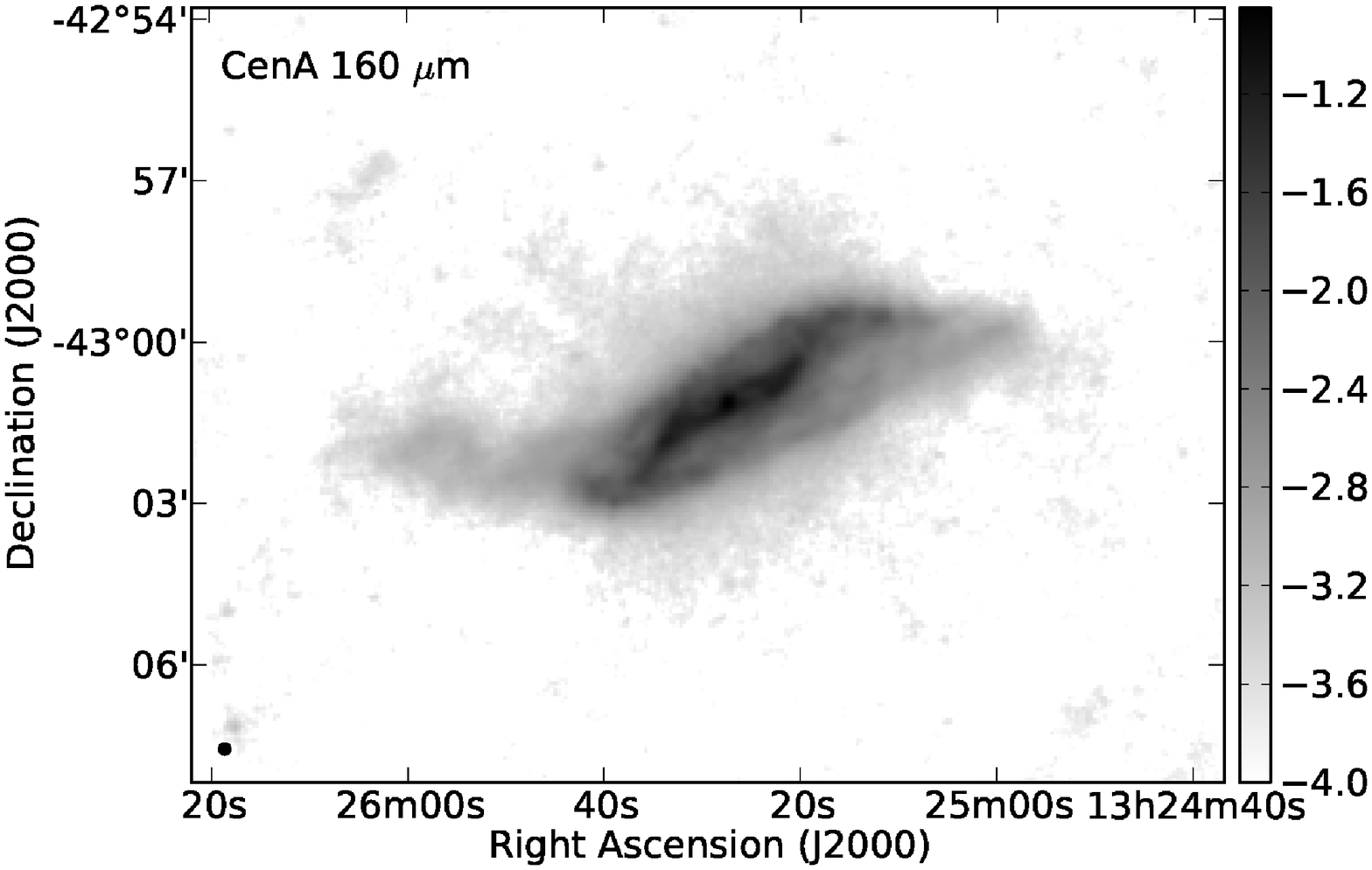}
\includegraphics[width=7.75cm]{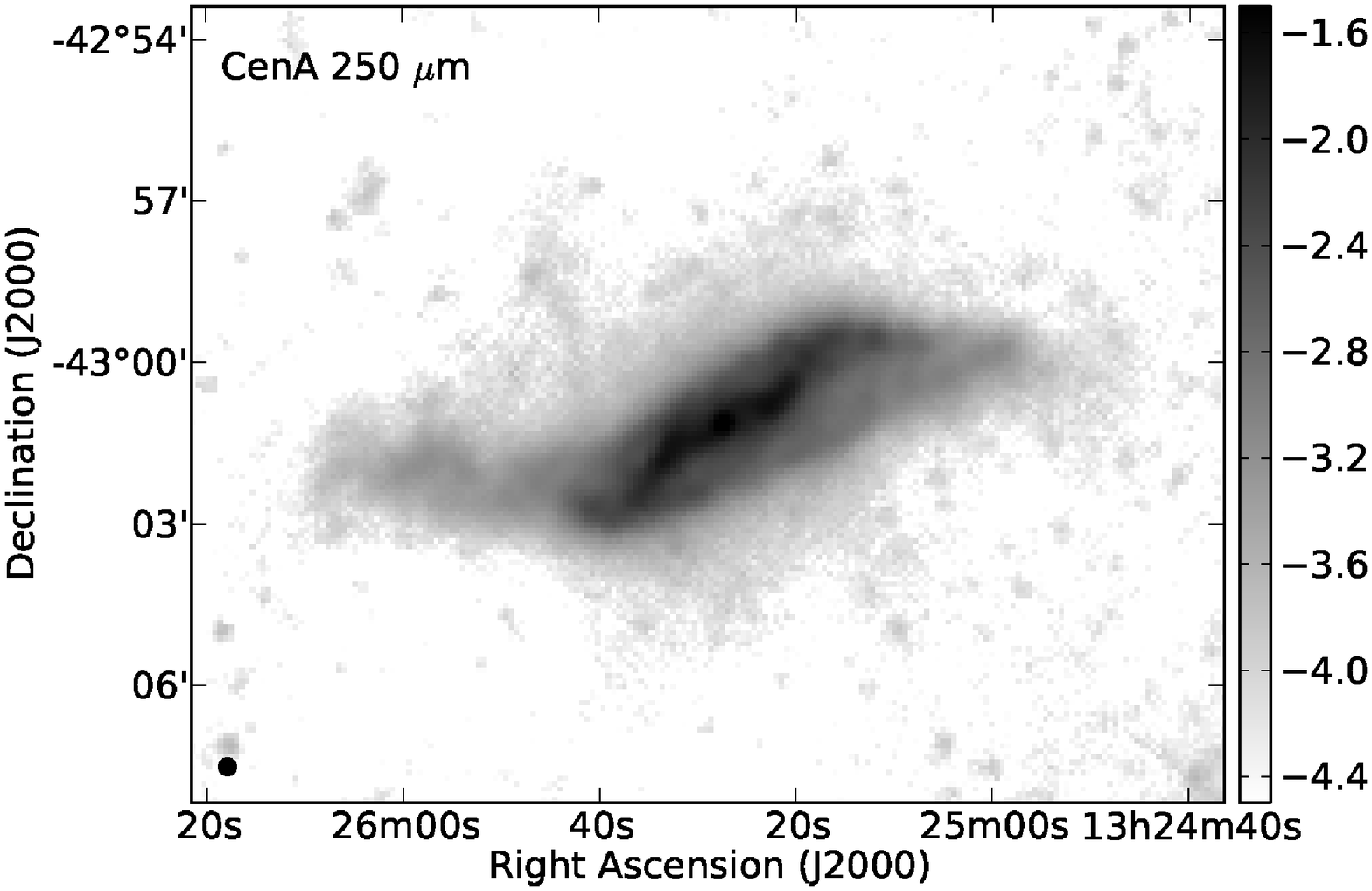}
\includegraphics[width=7.75cm]{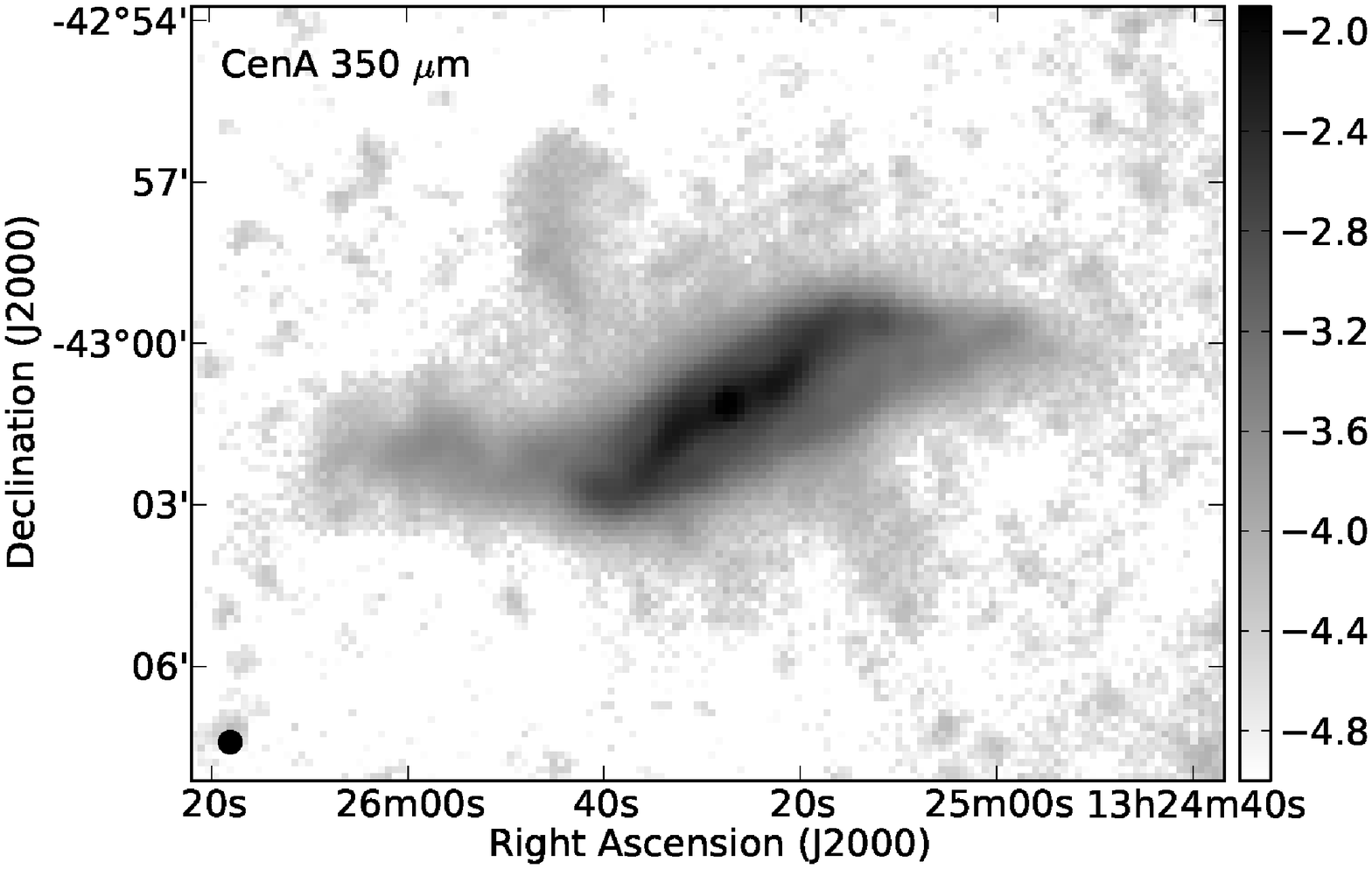}
\includegraphics[width=7.75cm]{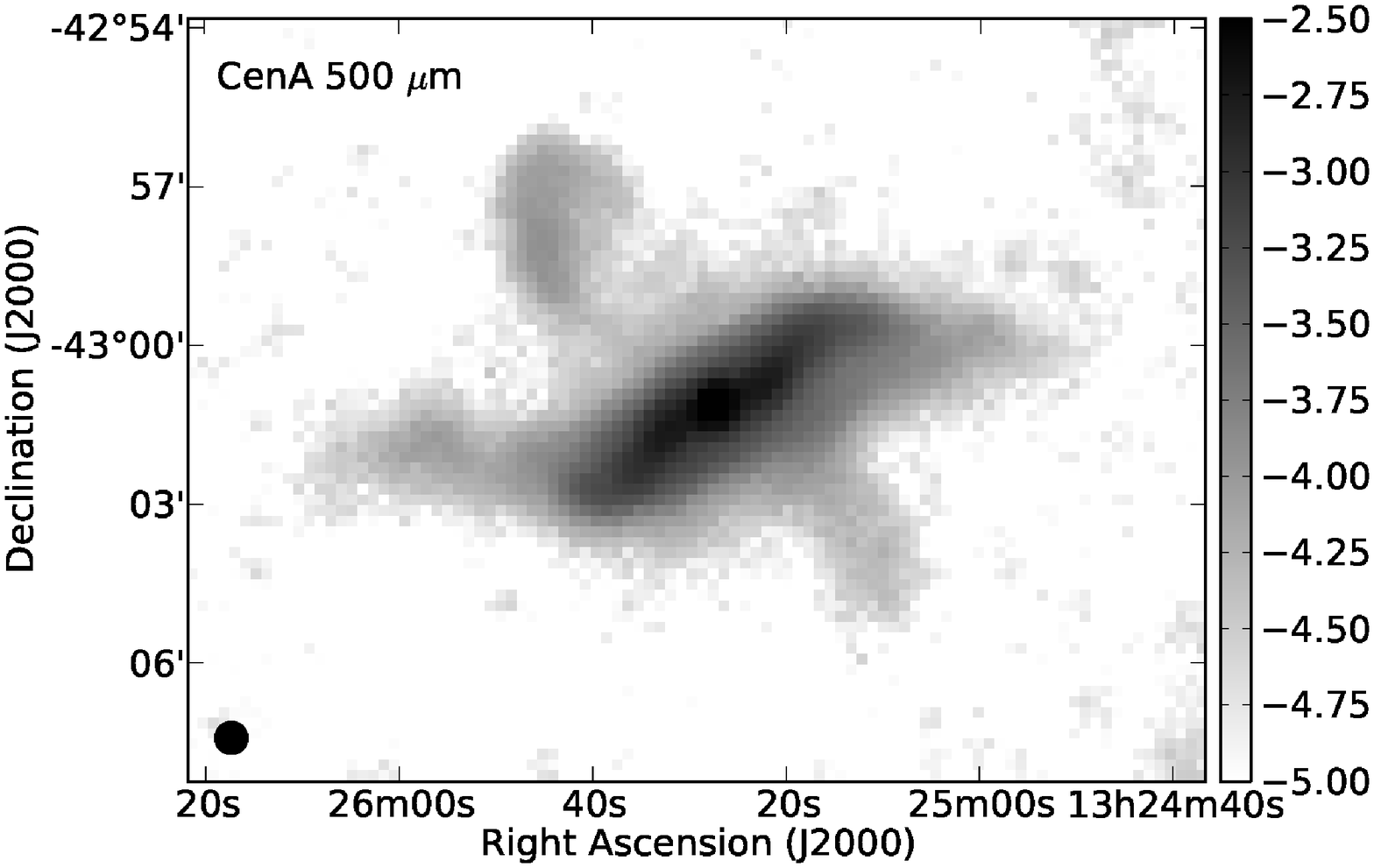}
\caption{The five \emph{Herschel} images at their native resolutions.  The colour scale units are log(Jy~arcsec$^{-2}$), and the beam size is in the lower left corner of each image.  Note that the radio jets become more prominent at 350 and 500~$\mu$m due to synchrotron radiation.}
\label{fig:Herschel_maps}
\end{figure*}

The SPIRE observations were done in large scan-map mode with the nominal scan speed of 30~arcsec~s$^{-1}$ and the cross-scanning method.  Data reduction was carried out in HIPE using a pipeline that has been altered from the standard one, in which the custom algorithm for temperature drift corrections, \emph{BRIght Galaxy ADaptive Element} (\textsc{BriGAdE}), was used in lieu of the standard algorithm.  A description of the data processing including how \textsc{BriGAdE} applies this correction is given in \citet{Auld_2011_submit}, \citet{Smith_2012_in_press} and Smith et al. (in prep).  The final maps were made using HIPE's map-making tool, and are in units Jy~beam$^{-1}$ with pixel scales of 6, 8 and 12~arcsec at 250, 350 and 500~$\mu$m, respectively.  However, we converted the units from Jy~beam$^{-1}$ to Jy~pix$^{-1}$ for consistency with the PACS images using beam areas for this conversion of 426, 771 and 1626~arcsec$^{2}$ for the 250, 350 and 500~$\mu$m images, respectively \citep{SOM_2010}.  We calibrated the flux to correct for the extended nature of Cen~A, using multiplicative K4 correction factors of 0.9828, 0.9834 and 0.9710 for the 250, 350 and 500~$\mu$m, respectively \citep{SOM_2010}.  Lastly, we applied a colour correction to each image using the same method as for the PACS images (see Table~\ref{table: herschel_char} for specific values), as well as a background subtraction. The five \emph{Herschel} images are presented in Figure~\ref{fig:Herschel_maps} in their original resolutions.

For analysis purposes, we have convolved the 250 and 350~$\mu$m images with a Gaussian beam to match the 36~arcsec resolution of the 500~$\mu$m image.  The point-spread function of the SPIRE beam is well-represented by a Gaussian function down to about three percent of the peak in all three wavebands \citep{2010A&A...518L...3G}, justifying our use of a Gaussian kernel.  Next, these maps were regridded such that they all had the pixel scale of the 500~$\mu$m image, 12~arcsec.

\subsubsection{Noise and Uncertainties}\label{subsec:Noise}
There are three sources of noise and uncertainty that we take into account for the \emph{Herschel} images.  The dominant uncertainty for all images is the calibration uncertainty.  For the PACS 70 and 160~$\mu$m images, the calibration uncertainties are 3 and 5~percent, respectively \citep{PACS_OM}.  For the 250, 350 and 500~$\mu$m images, the calibration uncertainties are taken to be 7~percent \citep{SOM_2010}.  There is also an underlying measurement uncertainty associated with the flux values in each pixel, and a map with these uncertainties is produced as part of the general reduction.  Finally, there is a contribution to the error in each pixel from the background subtraction.  To calculate this noise, we first find the mean value within a number of apertures overlaid on the convolved and regridded images, just as we did for the background subtraction.  Then we find the average mean across all apertures (at each wavelength), and then determine the standard error, given by the standard deviation of the individual aperture mean values divided by the square root of the number of apertures.  Note that we have not accounted for the confusion noise in individual pixels in the SPIRE maps \citep[see ][]{2010A&A...518L...5N}; however, we expect it to be trivial except perhaps in the extreme outer disk, and the global average of the confusion noise has been removed with the background subtraction.  A summary of the pixel errors from background subtraction and calibration uncertainties is presented in Table~\ref{table: herschel_char}.

\subsection{JCMT Observations}\label{subsec:JCMT}
 The $^{12}$CO~$J = 3-2$ transition at a rest frequency of 345.79~GHz was mapped using the HARP-B instrument on the JCMT, located on Mauna Kea in Hawaii.  These observations were taken over the nights of 2010 January 24--26 as part of project M09BC05 (P.I.: Tara Parkin), with a telescope beam size of 14.5~arcsec.  We combined nine overlapping jiggle-maps (each with an integration time of 50~s) instead of creating a raster map due to the low elevation of Cen~A from the location of the JCMT.  The maps were observed using beam-switching with a chop throw of 150~arcsec perpendicular to the major axis of Cen~A.  Each jiggle-map covers a $2 \mathrm{\,arcmin} \times 2 \mathrm{\,arcmin}$ area on the sky, and our final map covers an area of $10 \mathrm{\,arcmin} \times 2 \mathrm{\,arcmin}$, to observe the dust lane out to $D_{25}/2$, which is half the distance from the centre of the galaxy to where the optical magnitude falls to 25.  The backend receiver, the Auto-Correlation Spectrometer Imaging System (ACSIS), was set to a bandwidth of 1~GHz with 2048 channels, giving us a resolution of 488 kHz or 0.43~km~s$^{-1}$.

The data were reduced using the \textsc{Starlink}\footnote{The \textsc{Starlink} package is available for download at http://starlink.jach.hawaii.edu.} software package \citep{2008ASPC..394..650C}, maintained by the Joint Astronomy Centre.  We first flagged the data for any bad pixels in the raw data, and then built a cube (two spatial dimensions and one in frequency) with 7.5~arcsec pixels combining all of our observations.  Next, we masked out detections of the $^{12}$CO~$J = 3-2$ line in our cube using an automated fitting routine, then fit a third order polynomial to the baseline in order to remove it from the cube.  We then binned this cube to a resolution of 20~km~s$^{-1}$ to measure the average rms in the line free regions, which we determined to be $T_{\mathrm{A}}^{\ast} = 16$~mK.

The first three moment maps (integrated intensity, integrated velocity and velocity dispersion) were created using the routine `findclump' (part of \textsc{Starlink}), which looks for detections of greater than $2\sigma$.  We first convolve the unbinned cube with a Gaussian beam to match the 36\,arcsec resolution of the \emph{Herschel} SPIRE 500~$\mu$m map.  Next, we determine the average rms within the line free regions of the spectrum and expand the two-dimensional noise map into a cube.  Then we create a signal-to-noise cube by dividing the convolved cube by the noise cube and apply `findclump' to it.  This method allows us to make detections of low-level emission better than just applying `findclump' to our image cube.  The routine creates a mask of the detections in a smoothed version of the cube and then multiplies this mask by the signal-to-noise cube to obtain a cube containing only detections above several different $\sigma$ cutoffs.  Then this cube is collapsed along the velocity axis to create the three moment maps in terms of the signal-to-noise.  Finally, the integrated intensity-to-noise map is multiplied by the two-dimensional noise map to obtain the true integrated intensity map.  For a more detailed description of our reduction method see \citet{2010ApJ...714..571W}.

\begin{figure*}
\includegraphics[width=15cm]{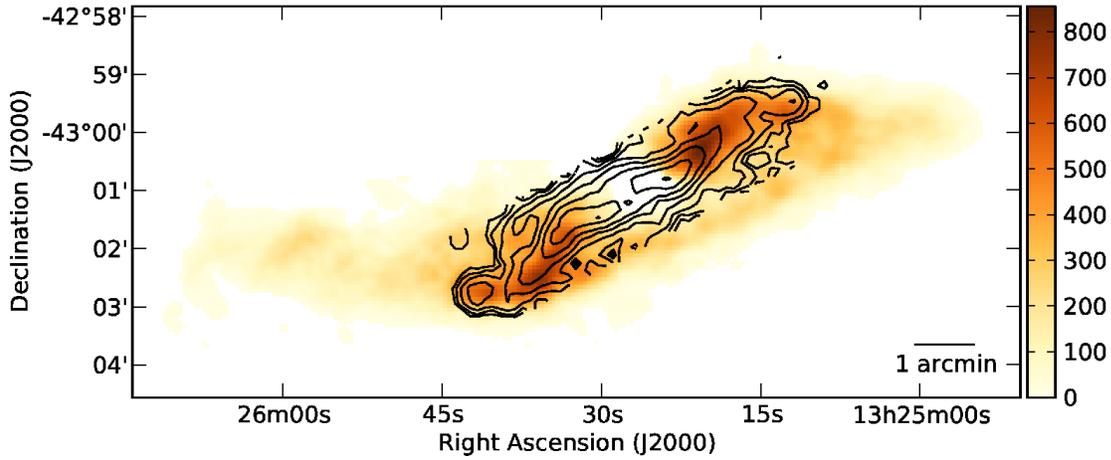}
\caption{H\textsc{i} map of the disk region of Cen~A in units of $10^{19}$~cm$^{-2}$ from \citet{2010A&A...515A..67S} (colourscale), with a beam of 19~arcsec.  JCMT CO~$J=3-2$ contours with a resolution of 14.5~arcsec are overlaid in black.  CO contour levels are 0.4, 0.8, 1.6, 3.2~K~km~s$^{-1}$\ldots25.6 and 38.4~K~km~s$^{-1}$, where the temperature units are $T_{\mathrm{A}}$*.}
\label{fig:co_map}
\end{figure*}

To evaluate the uncertainty in the integrated intensity map, we make use of the following equation:
\begin{equation}
 \Delta I = \left[ \Delta v \sigma \sqrt{N_{\mathrm{line}}} \right] \sqrt{1 + \frac{N_{\mathrm{line}}}{N_{\mathrm{base}}}},
\end{equation}
where $\Delta v$ is the width of each channel in km~s$^{-1}$, $\sigma$ is the noise in units of Kelvin, $N_{\mathrm{line}}$ is the total number of channels within the spectral line itself, and $N_{\mathrm{base}}$ is the number of channels used for fitting the baseline.  Several calibration sources measured during these observations were compared to standard spectra, and from these we estimate the calibration uncertainty to be 10~percent.  The integrated intensity map is presented as contours in Figure~\ref{fig:co_map}.

\subsection{Ancillary Data}\label{subsec:Ancillary}
The H\,\textsc{i} map of Cen~A was kindly provided to us by Tom Oosterloo and previously published in \citet{2010A&A...515A..67S}.  This map has a resolution of 19\,arcsec and units of $10^{19}$~cm$^{-2}$, with a pixel scale of 4\,arcsec.  We smoothed the image to a resolution of 36\,arcsec using a Gaussian beam to match the resolution of the SPIRE 500~$\mu$m image, and regridded the image to match the 12\,arcsec pixel size.

We also have a map of the radio continuum at 1.425\,GHz, originally published as part of the \emph{IRAS} Bright Galaxy Atlas \citep{1996ApJS..103...81C}, and retrieved from the NASA Extragalactic Database (NED).  This map allows us to trace the jets extending out from the central AGN.  We convolved this radio image with a Gaussian beam and then regridded it to match the SPIRE 500~$\mu$m image.

\section{Results}\label{sec:results}
\subsection{SED Fitting}\label{subsec:sed}
To determine the dust temperatures within the disk of Cen~A, we fit the far-infrared and submm part of the spectral energy distribution (SED; 70 -- 500~$\mu$m) with a simple modified blackbody with $\beta = 2$,
\begin{equation}\label{eqn:blackbody}
 I(\nu,T) = C \times \nu^{\beta} \times B(\nu,T),
\end{equation}
where $C$ is a scaling constant to match the model to our observed fluxes.  The parameter $\beta$ originates from the dust emissivity function, $\kappa_{\nu} = \kappa_{0}(\nu/\nu_{0})^{\beta}$.

To constrain our modified blackbody fits, we created uncertainty maps for each wavelength that are a combination of the measurement uncertainty in the flux value of each pixel, the background noise in each pixel, and the calibration uncertainties for both PACS and SPIRE (see Section~\ref{subsec:Noise} for details), added in quadrature.  The value in each pixel of the resulting map is taken to be the total uncertainty for the equivalent pixel at a given wavelength.

\begin{figure*}
 \includegraphics[width=7.5cm]{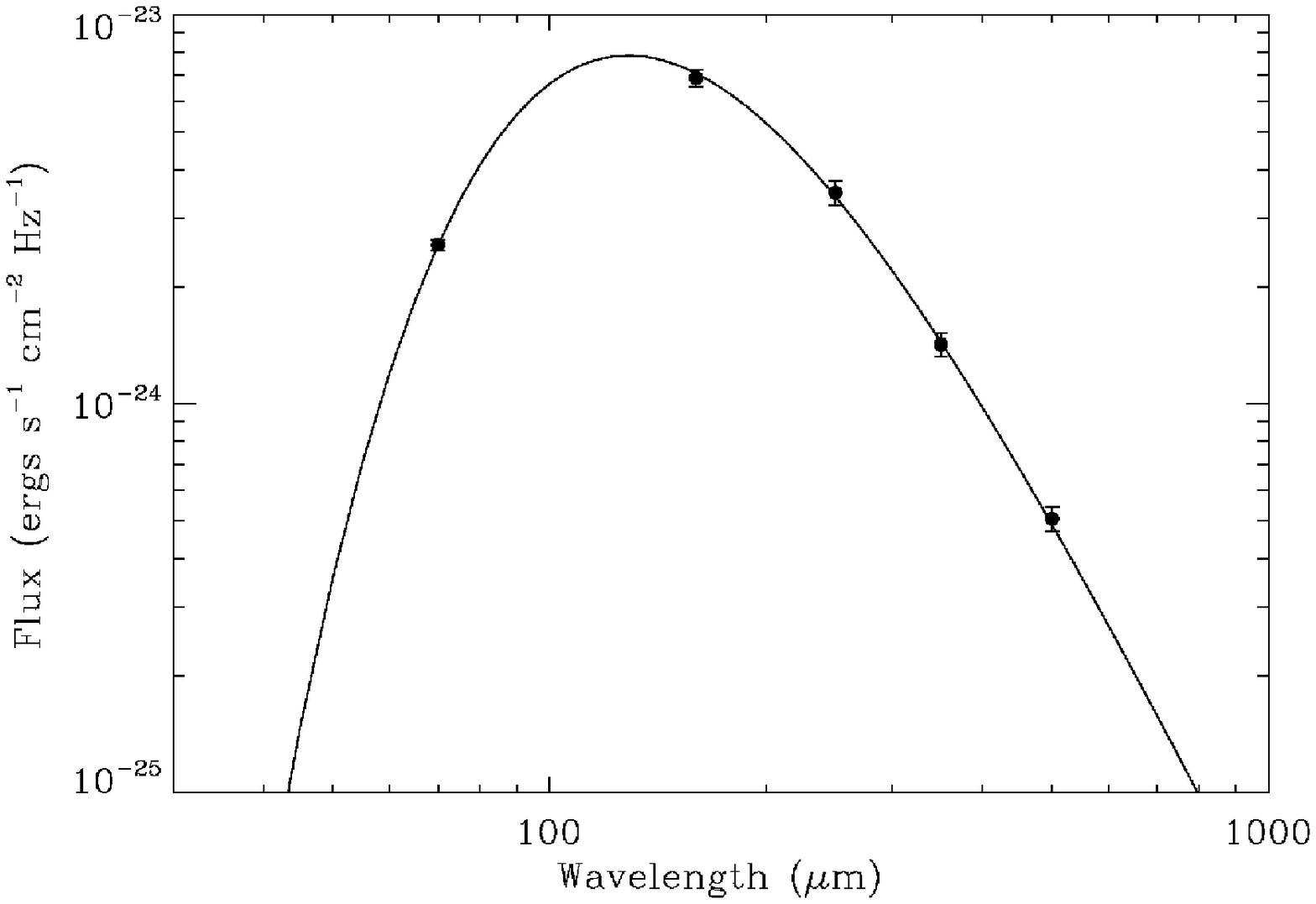}
 \includegraphics[width=7.5cm]{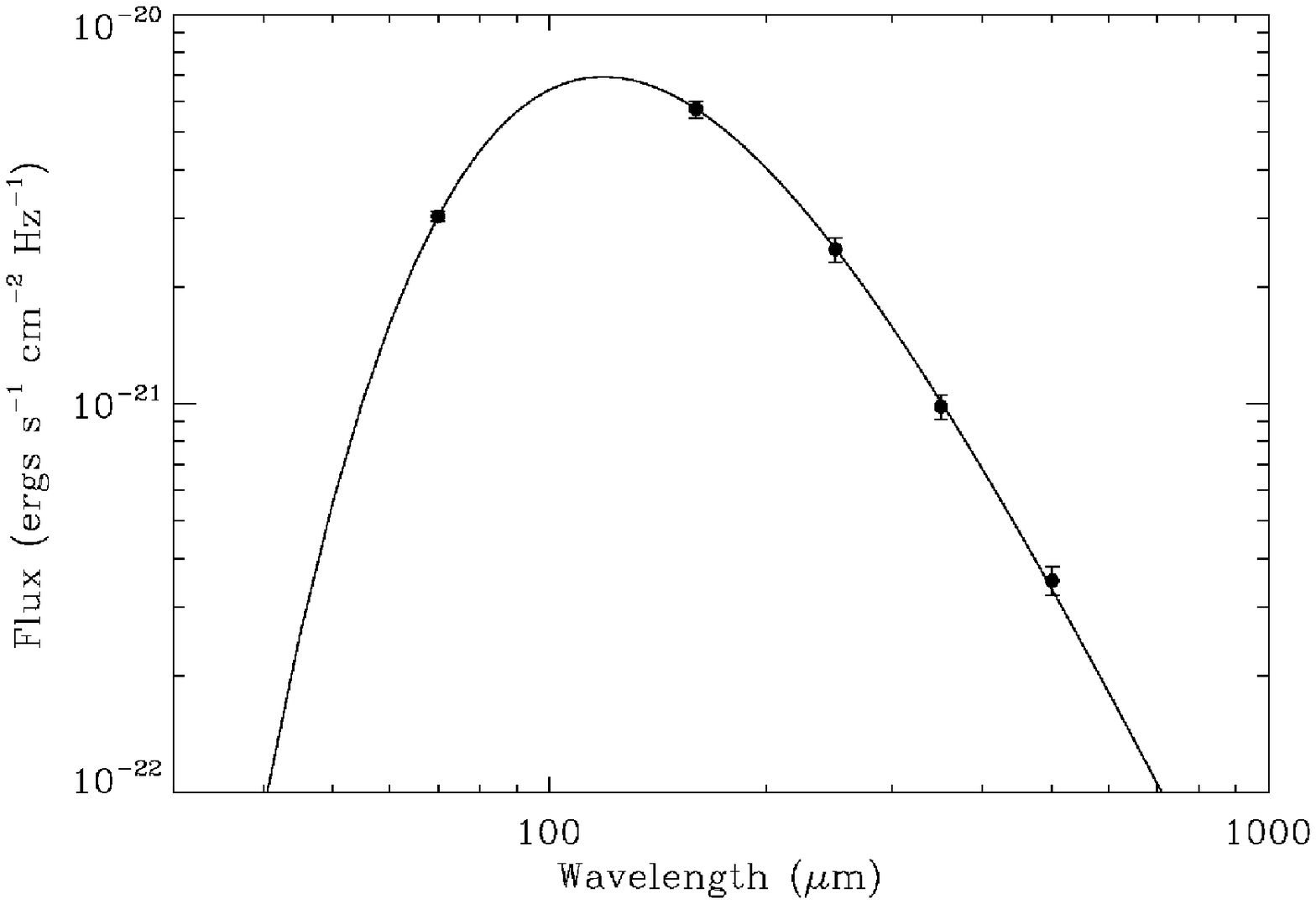}
\caption{\textit{left}: The SED for a typical pixel (pixel location shown as a cross in Figure~\ref{fig:temp}).  The best-fitting modified blackbody is shown as a solid line, while our measured flux values are represented by the data points. \textit{right}: A global SED for the entire disk, consisting of all pixels contributing to the total dust mass (see Section~\ref{subsec:dust}).}
\label{fig:sed}
\end{figure*}

\begin{figure*}
\includegraphics[width=15cm]{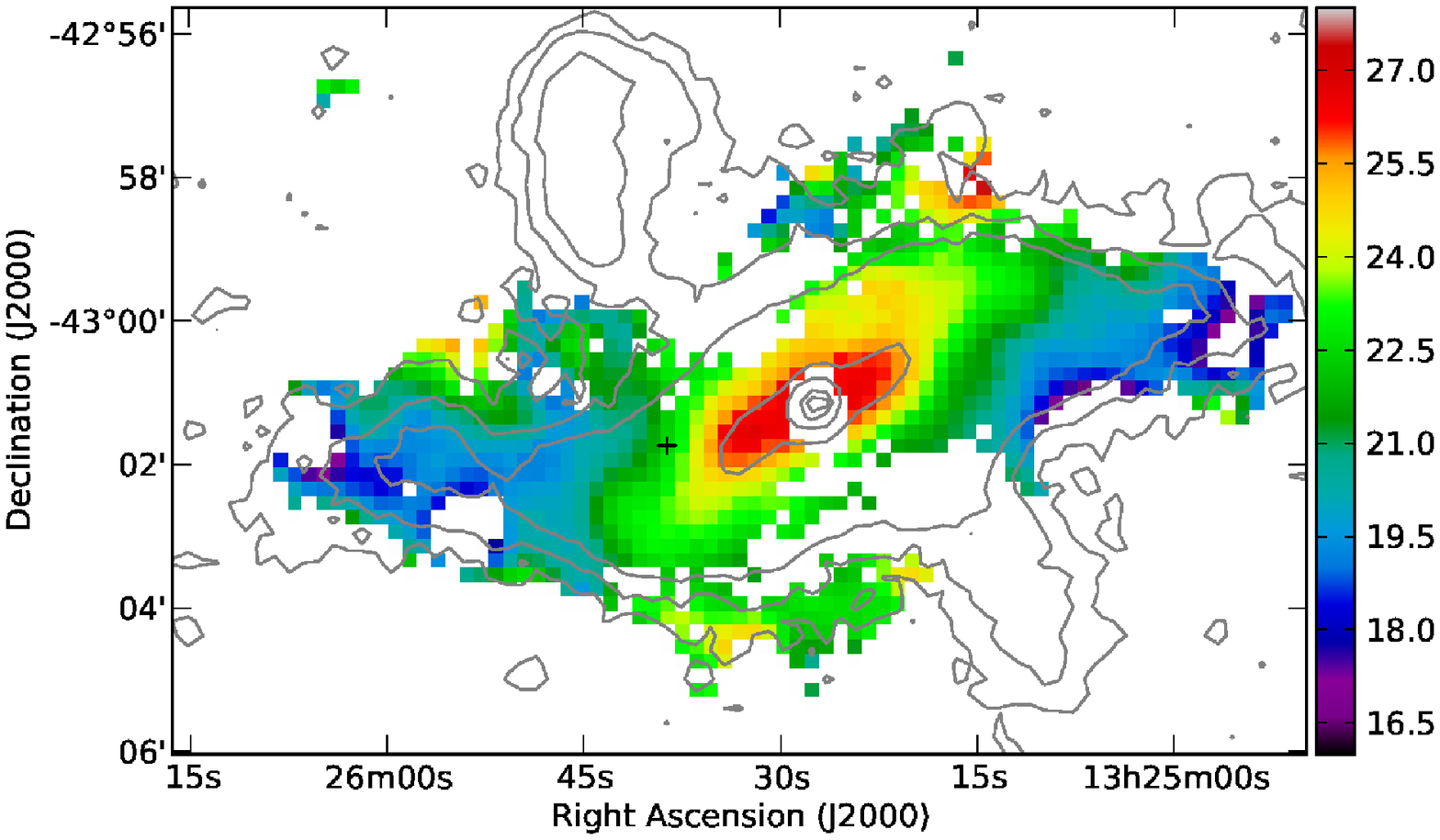}
\caption{Temperature map of the dust disk of Cen~A (colour scale), in units of degrees Kelvin.  Contours of the 500~$\mu$m SPIRE map are shown in grey for reference.  The contours for the 500~$\mu$m image are 0.0139, 0.0347, 0.0694, 0.347, 0.694, 1.39, 2.78, 4.17 and 4.86~mJy~arcsec$^{-2}$.  The black cross is the location of the pixel for which the best-fitting SED is shown in Figure~\ref{fig:sed}. Note that the central region does not have any reasonable fits due to the non-thermal emission present in the 350 and 500~$\mu$m maps.}
\label{fig:temp}
\end{figure*}

For each pixel in the images that has a S/N~$\ge 10$ at all five \emph{Herschel} wavelengths, we fit Equation~\ref{eqn:blackbody} to our flux measurements to determine the best-fitting temperature $T$ and constant $C$ for that pixel.  An example of a SED fit for a typical pixel is shown in Figure~\ref{fig:sed}, and the pixel location is shown with a cross in Figure~\ref{fig:temp}.  We also show the global SED that was obtained by summing the flux at each wavelength within all of the pixels with good fits in Figure~\ref{fig:sed}, and fitting the resulting totals.  Note that we tested the effect of removing the 70~$\mu$m data point from our SED fit to ensure it was not forcing the curve to peak at shorter wavelengths, and that it was not from a separate thermal component of the dust \citep{2010A&A...518L..51S, 2010A&A...518L..65B, Bendo_2011_submit}.  We found that the 70~$\mu$m flux falls on the best-fit curves within uncertainties for these tests and thus we use all five wavelengths in our fitting routine.  We then create a map of the dust temperature where pixels with poor chi-squared fits at the 95~percent confidence level have been masked out.  In general the pixels with poor fits coincide with those where non-thermal radiation is making a significant contribution to the flux at 500~$\mu$m. The temperature map is presented in Figure~\ref{fig:temp}, with 500~$\mu$m contours overlaid to point out the location of the non-thermal emission.  The temperature is about 30~K near the central AGN region, and then smoothly falls off to about 20~K in the outer disk, and the average temperature throughout the entire region is approximately ($21.8 \pm 0.3$)~K.

We have also tried allowing both $\beta$ and temperature to vary in Equation~\ref{eqn:blackbody}, to investigate if the model fits to our data improve.  While there is some small variation in $\beta$ throughout the disk, these deviations are small and the average value of $\beta$ is $2.07 \pm 0.07$.  Furthermore, there are no obvious trends in the variation of $\beta$ with radius.  Thus, the results of our test are consistent with our one parameter fit, giving us confidence in our fixed $\beta$ fits.

We evaluated the uncertainty on the dust temperature by implementing a monte carlo algorithm on our modified blackbody fitting routine.  For each wavelength, an artificial flux value lying within the uncertainty bounds of our observed measured flux value is generated randomly using a Gaussian number generator.  Then we run the modified blackbody fitting routine on the simulated dataset and extract the temperature and constant for the best fitting function corresponding to that dataset.  This process of creating artificial data and then fitting a function to them to determine the temperature is repeated 800 times for each pixel.  We take the standard deviation of the temperature distribution for each pixel to be the uncertainty in the temperature.  This uncertainty varies throughout the map, with an average of about 1~percent per pixel.  The highest uncertainties are $5$~percent around the very edge of our map, where the signal to noise is lower for the flux maps, and we note that these uncertainties exclude the uncertainties arising from our assumed value for $\beta$ and thus emissivity.

\subsection{Dust Mass}\label{subsec:dust}
The equation we use to evaluate the dust mass at a specific frequency is
\begin{equation}\label{eqn:dust_mass}
 M_{\rmn{dust}} = \frac{S_{\nu}D^{2}}{\kappa_{\nu} B(\nu,T)},
\end{equation}
when the monochromatic flux is optically thin.  In this expression, $S_{\nu}$ is the flux from the source, $D$ is the distance to the source, $B(\nu,T)$ is the equation for a blackbody and $\kappa_{\nu}$ is the dust emissivity.  Here we adopt a value for the dust emissivity at 250~$\mu$m, $\kappa_{250}$, of 3.98~cm$^{2}$~g$^{-1}$ from the dust model of \citet{2003ARA&A..41..241D}\footnote{These data are also available at \newline http://www.astro.princeton.edu/~draine/dust/dustmix.html}.  Substituting $\kappa_{250}$ into Equation~\ref{eqn:dust_mass}, the dust mass becomes
\begin{equation}\label{eqn:spire_dust_mass}
 M_{\rmn{dust}} = 5.258 \times 10^{3}S_{250}\left(e^{\frac{57.58~\mathrm{K}}{T}}-1\right) ~\mathrm{M}_{\odot}.
\end{equation}
We checked that the SED of Cen~A is optically thin by assuming that the pixel with the highest flux at 70~$\mu$m is a worst-case scenario in terms of optical depth $\tau$, and measured the flux in the same pixel at 250~$\mu$m.  Using the results from our SED fitting we calculate the dust mass within that pixel using Equation~\ref{eqn:spire_dust_mass}, and convert the dust mass to a dust mass surface density $\Sigma$ using the area of one 12~arcsec pixel. The optical depth at 250~$\mu$m can be calculated as $\tau_{250} = \Sigma \kappa_{250}$.  Using this method we find that $\tau_{250} \ll 1$, thus we can assume the galaxy is optically thin at 250~$\mu$m.  This method can be repeated for the other wavelengths as well.

\begin{figure*}
\includegraphics[width=15cm]{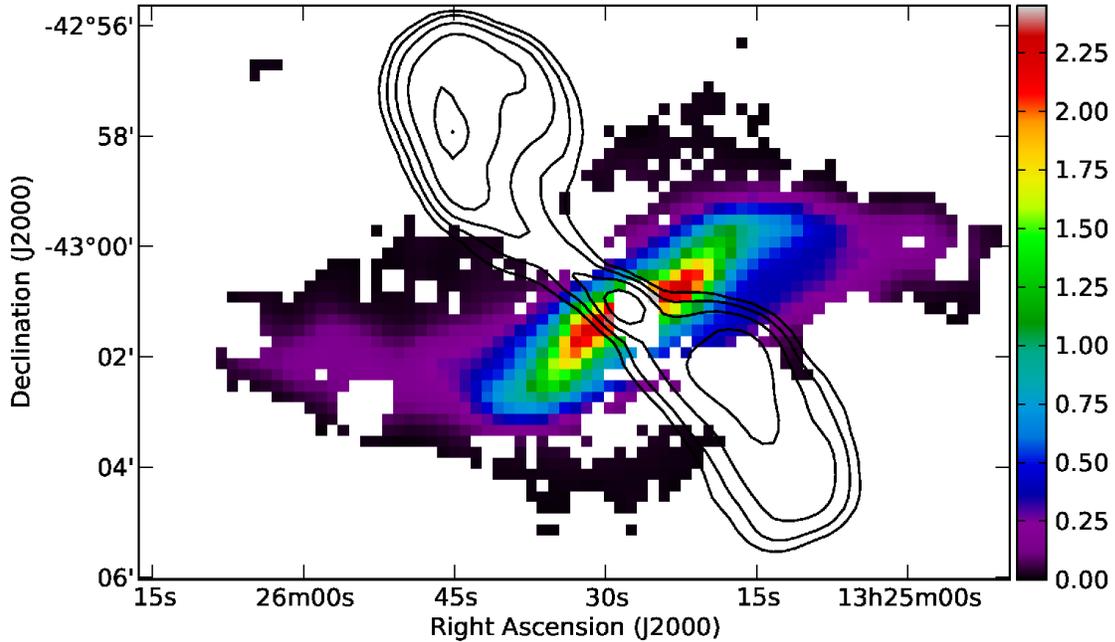}
\caption{Dust mass map of Centaurus A.  Units of the colour scale are M$_{\odot}$~pc$^{-2}$.  Contours of the 1.425~GHz radio continuum are overlaid in black to show the regions where synchrotron radiation is present.}
\label{fig:dust_mass}
\end{figure*}

Substituting our temperature and the SPIRE 250~$\mu$m fluxes returned by our model SED fits into Equation~\ref{eqn:spire_dust_mass}, we obtain a  pixel-by-pixel map of the dust mass, which is shown in Figure~\ref{fig:dust_mass}.  The dust mass distribution peaks toward the centre, and falls off in the outer regions.  Uncertainties range from about 3 to 30~percent with the lower uncertainties corresponding to pixels closer to the centre of the disk.  The typical uncertainty in each pixel is $\sim 5$~percent.  Summing up the individual dust masses within each pixel leads to a total dust mass in the disk of $(1.59 \pm 0.05) \times 10^{7}$~M$_{\odot}$.  As a check, we also evaluated the total dust mass using the global SED fit results, and obtain a total of $1.47 \times 10^{7}$~M$_{\odot}$.

\subsection{Gas Mass}\label{subsec:gas}
We convert our CO~$J=3-2$ integrated intensity map to a column density of molecular hydrogen using the following equation:
\begin{equation}\label{eqn:CO_col_density}
N_{\mathrm{H}_{2}} = \frac{\mathrm{X}_{\mathrm{CO}} I_{\mathrm{CO}(3-2)}}{\eta_{\mathrm{mb}}} \left(\frac{I_{\mathrm{CO}(3-2)}}{I_{\mathrm{CO}(1-0)}}\right)^{-1},
\end{equation}
where X$_{\mathrm{CO}}$ is the CO($J=1-0$)-H$_{2}$ conversion factor, $I_{\mathrm{CO}(3-2)}$ is the integrated intensity of the CO~$J=3-2$ emission, $\eta_{\mathrm{mb}}=0.6$ is the conversion factor from an antenna temperature $T_{\mathrm{A}}$* to a main beam temperature $T_{\mathrm{mb}}$ for the JCMT, and the factor within the brackets is the ratio of the CO~$J=3-2$ line intensity to the CO~$J=1-0$ line intensity.  Despite the additional uncertainty introduced by this line intensity ratio, we choose to use our CO~$J=3-2$ map in this analysis because of its better intrinsic resolution (14.5~arcsec) and sensitivity compared to the CO~$J=1-0$ of \citet{1990ApJ...363..451E}.  We assume a X$_{\mathrm{CO}}$ factor of $(2 \pm 1) \times 10^{20}$~cm$^{-2}$~(K~km~s$^{-1}$)$^{-1}$, typical for the Milky Way \citep{1988A&A...207....1S}, and a CO~$J=3-2$/CO~$J=1-0$ ratio of 0.3, which is a suitable ratio for the diffuse ISM as found by the JCMT Nearby Galaxies Legacy Survey \citep[NGLS;][]{2009ApJ...693.1736W}.  We note here that \citet{2009ApJ...693.1736W} and other groups from the NGLS have seen ratio variations within other galaxies by up to a factor of 2-3, which may increase our molecular gas mass uncertainties (see Section~\ref{sec:discuss} for further discussion).

Next, we convert our column density to a molecular gas mass map via
\begin{equation}\label{eqn:mol_gas_mass}
M_{\mathrm{H}_{2}} = A_{\mathrm{pix}} N_{\mathrm{H}_{2}} m_{\mathrm{H}_{2}}
\end{equation}
with $A_{\mathrm{pix}}$ the area of one pixel (12~arcsec~$\times$~12~arcsec) at the distance of Cen~A, and $m_{\mathrm{H}_{2}}$ the mass of one molecular hydrogen atom.  The total mass of molecular hydrogen, over the entire coverage of our map, is $(1.42 \pm 0.15) \times 10^{9}$~M$_{\odot}$, where we have excluded uncertainty contributions from the assumed CO~$J=3-2$/CO~$J=1-0$ ratio.

\begin{figure*}
\includegraphics[width=15cm]{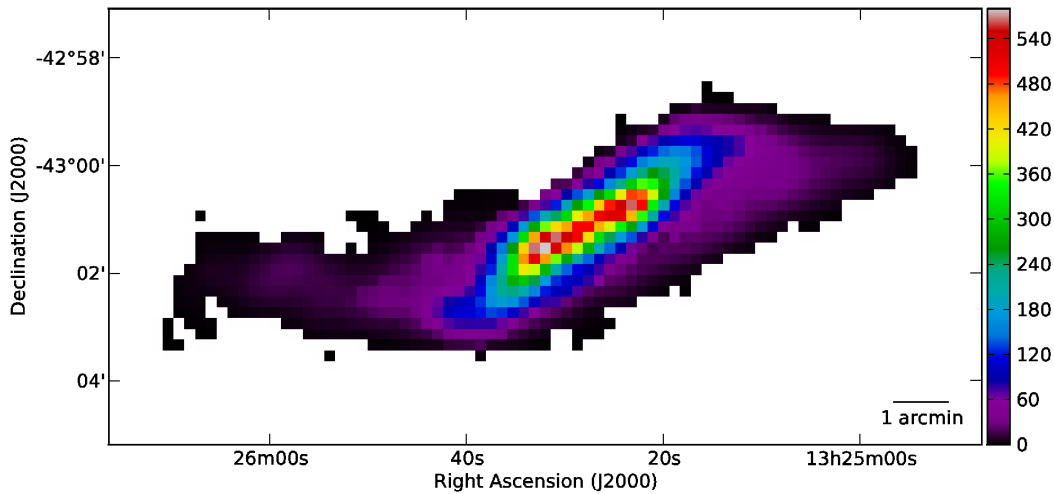}
\caption{The total gas mass distribution of the disk of Cen~A.  This map includes both the H\textsc{i} mass distribution, as well as the H$_{2}$ mass, where we have coverage for it, and has been multiplied by 1.36 to account for helium as well.  Units for the colour scale are M$_{\odot}$~pc$^{-2}$.}
\label{fig:gas}
\end{figure*}

Following the same method we also convert the units of the H\textsc{i} map (Figure~\ref{fig:co_map}) to units of mass.  Then the molecular gas mass and atomic gas mass maps are combined in such a way that in regions where we have coverage for both H\textsc{i} and H$_{2}$, the total gas mass is a sum of the two, and for the remaining regions the mass is from H\textsc{i} or H$_{2}$ alone.  We then multiply the resulting map by 1.36 to obtain the total gas mass, including helium.  Summing over all pixels, we find a total gas mass of $(2.7 \pm 0.2)\times 10^{9}$~M$_{\odot}$.  Our map of the total gas is presented in Figure~\ref{fig:gas}, and covers about half of the area for which we have dust mass measurements.

\subsection{The Gas-to-Dust Mass Ratio}\label{subsec:gas2dust}
\begin{figure*}
\includegraphics[width=15cm]{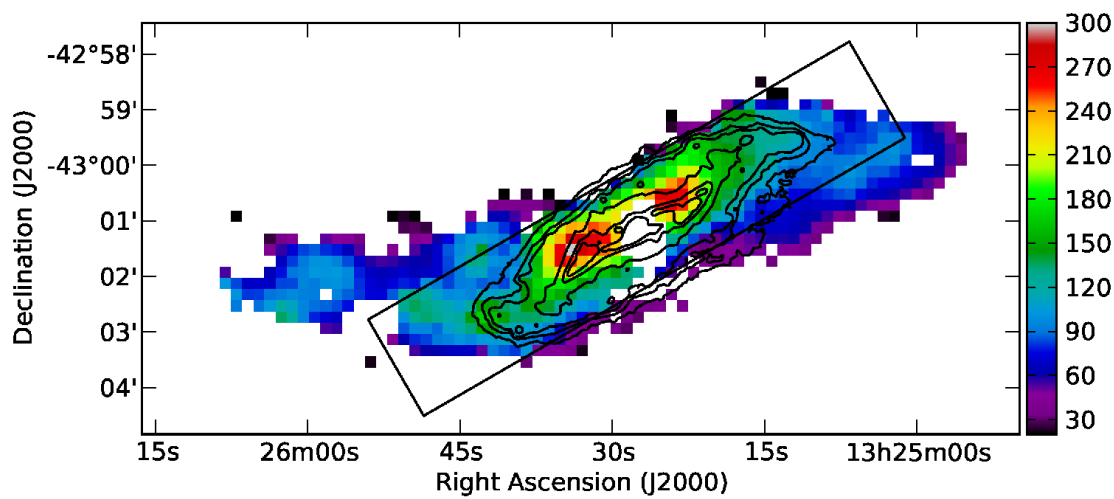}
\caption{Gas-to-dust mass ratio of Centaurus~A (colourscale), with PACS 70~$\mu$m contours (at original resolution) overlaid in black.  Contour levels are 0.75, 1.5, 2.5, 7.5, 20, and 37.5~mJy~arcsec$^{-2}$.  The box shown in black highlights the region in the disk for which the gas contribution is from H\textsc{i} and H$_{2}$, and is 10~arcmin~$\times$~2~arcmin in size.}
\label{fig:g2d}
\end{figure*}

\begin{figure}
\includegraphics[width=8cm]{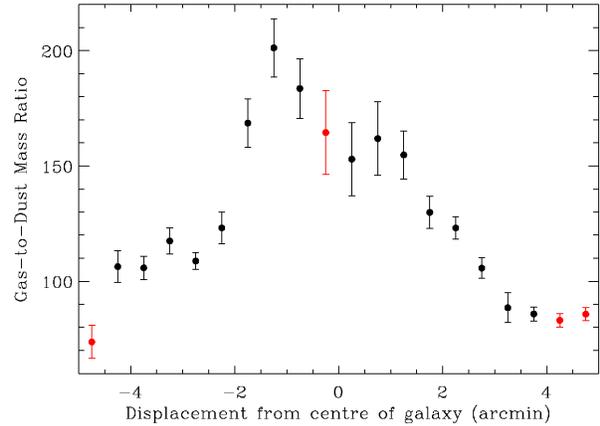}
\caption{A radial plot of the average gas-to-dust mass ratio along the major axis of Cen~A with a position angle of 30~deg north of west.  We measured the average gas-to-dust ratio value within 20 bins along this position-angle, each measuring 2~arcmin~$\times$~0.5~arcmin in size.  The red data points represent the bins in which we had at most 10 good pixels out of a possible 25, and therefore are not as well sampled as those shown in black.}
\label{fig:g2d_radial}
\end{figure}

We present the gas-to-dust mass ratio in Figure~\ref{fig:g2d}, which covers all pixels for which we have both a gas mass and a dust mass.  The box outlined in black shows the region that has coverage both in H\textsc{i} and H$_{2}$, which is 10~arcmin~$\times$~2~arcmin in area.  As is seen in this figure, the majority of the disk has a gas-to-dust ratio similar to that of the Galaxy with an average value of $\mathbf{103 \pm 8}$; the Galactic value varies from about 120 \citep{2001ApJ...554..778L} to $\sim 160$ \citep{2004ApJS..152..211Z} to $\sim$180 \citep[][and references therein]{2007ApJ...663..866D}.  Interestingly, the gas-to-dust ratio increases to a significantly higher value of almost 300 in regions closest to the AGN.  Typical uncertainties in this ratio are about 10~percent.  We also present the average gas-to-dust ratio as a function of distance from the centre of Cen~A along a position angle of 30~deg north of west in Figure~\ref{fig:g2d_radial}. This plot emphasizes the radial trend of the gas-to-dust ratio along the disk. In other galaxies such as those from the SINGS survey, the gas-to-dust ratio varies from $\sim 136$ to $\sim 453$ \citep{2007ApJ...663..866D}.  We discuss the possible origin of this high central gas-to-dust mass ratio below.

\section{Discussion}\label{sec:discuss}
Our total dust mass of $(1.59 \pm 0.05) \times 10^{7}$~M$_{\odot}$ is comparable to other published numbers for Cen~A in the literature; however, we note that this value is likely a lower limit as regions where non-thermal emission contaminates the dust continuum had poor model fits.  Several publications have used two-component modified blackbody modelling combined with far-infrared and submm observations to derive a dust mass for Cen~A.  \citet{2002ApJ...565..131L} used a wavelength range of 60--850~$\mu$m for their SED fit and found a total mass of $2.6 \times 10^{6}$~M$_{\odot}$ (adjusting for our distance) within an elliptical region 120~arcsec~$\times$~450~arcsec in size, but excluding the unresolved core at the centre.  Since this ellipse covers a similar area to our observations, the discrepancy is likely due to an inadequate chop throw in the \citet{2002ApJ...565..131L} 850~$\mu$m dataset, leading to difficulties quantifiying the background.  Our observations show faint emission at 500~$\mu$m extending beyond the 120~arcsec chop throw the authors used, and oversubtracting the background will lead to a lower dust mass estimate.  \citet{2008A&A...490...77W} studied the submm continuum using LABOCA 870~$\mu$m data and combining these data with fluxes spanning 25--870~$\mu$m they found a higher mass of $2.2 \times 10^{7}$~M$_{\odot}$.  They subtract the flux contribution of the unresolved core from the total flux prior to estimating the dust, but do caution that they may not have removed all of the non-thermal emission arising from the radio jet.

The gas mass we have obtained from our H$_{2}$ map alone is $(1.42 \pm 0.15) \times 10^{9}$~M$_{\odot}$, including the central region which we masked out for our dust fitting.  \citet{1990ApJ...363..451E} determined a molecular gas mass of $1.9 \times 10^{8}$~M$_{\odot}$, after we adjust their result for our assumed X$_{\mathrm{CO}}$ factor and distance.  In addition, \citet{2010PASA...27..463M} suggests the total molecular gas mass is $4 \times 10^{8}$~M$_{\odot}$ based on observations of \citet{1990ApJ...363..451E} and \citet{1992ApJ...391..121Q}.  Our molecular gas mass is thus higher than those values previously published; this is likely due to the improved sensitivity of our observations.  The molecular gas mass is about twice the atomic mass of $4.9 \times 10^{8}$~M$_{\odot}$ \citep{2010A&A...515A..67S}.

The high gas-to-dust mass ratio towards the central AGN is a most interesting result.  Several studies of spiral galaxies have found that the gas-to-dust ratio is not constant throughout the disk, but rather decreases with decreasing distance from the centre of the galaxy \citep[e.g.][]{2009ApJ...701.1965M,2010MNRAS.402.1409B,2011arXiv1106.0618M}, in contrast to what we have found here.  We have made a rough estimate of the CO~$J=3-2$/CO~$J=1-0$ ratio throughout the central disk using the CO~$J=1-0$ map of \citet{1990ApJ...363..451E} to check if our assumption of 0.3 for this ratio is valid in the centremost regions of the disk.  Our calculations show an average value of 0.35, and more importantly, they show no significant variation with radius.  Furthermore, looking at Figure~\ref{fig:g2d} we see that there is a smooth transition of the gas-to-dust mass ratio between the H$_{2}$-dominated region to the H\textsc{i}-dominated region in the outer disk, which also suggests CO excitation is not the primary cause of the high gas-to-dust ratio.

The X$_{\mathrm{CO}}$ factor is one variable that might affect the gas-to-dust ratio if it is not constant throughout the disk.  We have assumed the Galactic value for our calculations; however, studies have shown that this conversion factor can vary with metallicity \citep{1995ApJ...448L..97W, 1997A&A...328..471I, 2000mhs..conf..293I, 2000MNRAS.317..649B, 2003A&A...397...87I, 2004A&A...422L..47S, 2005A&A...438..855I, 2011ApJ...737...12L}.  If the metallicity decreases as a function of radius within Cen~A, we would require a larger conversion factor at larger radii, which in turn would increase the gas-to-dust ratio farther out in the disk, potentially bringing it up to a similar value as that in the centre.  Alternatively, the X$_{\mathrm{CO}}$ factor can be smaller by up to a factor of $\sim 4$ for starbursting galaxies \citep{1998ApJ...507..615D}.  This effect could reduce the gas-to-dust ratio in regions of strong star forming activity; however, the total far-infrared luminosity for Cen~A is $\le 10^{10}~L_{\odot}$, while the X$_{\mathrm{CO}}$ conversion factor is typically affected at luminosities greater than $\sim 10^{11}~L_{\odot}$ \citep[e.g.][]{1993ApJ...414L..13D, 1997ApJ...478..144S}.

One possibility is that dust is being destroyed by the jets extending out in either direction from the AGN.  It is plausible that some of the dust grains are being destroyed by the AGN either through dust sputtering or through shocks.  The total X-ray luminosity of the AGN of Cen~A is $8 \times 10^{42}$~erg~s$^{-1}$ \citep{2011ApJ...733...23R}, which is comparable in X-ray luminosity to the AGN found in M87 \citep{2006A&A...459..353W}.  It might be possible for these photons to deplete the dust in the surrounding ISM.  In support of this scenario, \citet{2010A&A...518L..53B} did not detect thermal dust emission in M87, which also contains a jet.

It is also possible that the dust is entrained in the jets of Cen~A.  \citet{2010A&A...518L..66R} found a dusty halo surrounding the starbursting galaxy M82 using \emph{Herschel} observations.  They considered entrainment by a galactic wind as a possible explanation even though they finally concluded that the origin was likely due to tidal interactions.  We cannot rule out entrainment by galactic winds as a possibility for Cen~A.  Unfortunately, we cannot probe close enough to the jets to further investigate this assumption due to our limited resolution and the non-thermal emission component.

Alternatively, a larger gas reservoir in the centre could also increase the gas-to-dust ratio.  This would require dust-poor gas falling onto the disk and being funneled toward the centre.  However, if this was the case, we would likely have found a high gas-to-dust throughout the disk and not concentrated toward the innermost part of the disk.  It also does not seem plausible for the gas to be migrating through the disk faster than the dust.

Interestingly, there is a large ring of material that is seen in the mid- and far-infrared close to the regions in which we see a high gas-to-dust ratio.  \citet{1999A&A...341..667M} first noticed the `S' shaped structure with ISOCAM observations at 7 and 15~$\mu$m, and postulated that this structure was a bar at the centre of the galaxy.  Later, observations \citep{2002ApJ...565..131L,2006ApJ...645.1092Q} supported a tilted ring scenario to explain the structure observed.  Our \emph{Herschel} observations at 70~$\mu$m also show emission from this ring (Figure~\ref{fig:Herschel_maps}), and in fact, the high gas-to-dust ratio appears to correspond with the 70~$\mu$m contours of the ring, as shown in Figure~\ref{fig:g2d}.  The models by \citet{2006ApJ...645.1092Q} suggest that there is a deficit in the dust distribution at radii less than about 50~arcsec ($\sim 920$~pc).  This is at a similar radius to our observed high gas-to-dust ratio.  A dust deficit throughout $r \le 50$~arcsec might seem less likely to be related to the AGN and is perhaps better related to star formation activity.  Observations probing closer to the nucleus are necessary to determine if the gas-to-dust ratio remains high or increases further as we approach the AGN.

\section{Conclusions}\label{sec:conclusions}
We have presented new observations of the dust continuum from the PACS and SPIRE instruments on board the \emph{Herchel Space Observatory} at 70, 160, 250, 350 and 500~$\mu$m.  In addition, we present new data tracing the CO~$J=3-2$ transition taken with the HARP-B instrument mounted on the JCMT.  We have used these data to probe the interstellar medium within the disk of Centaurus~A on a pixel-by-pixel basis. We observe the ring of emission at 70~$\mu$m previously reported at infrared and submm wavelengths, while our 500~$\mu$m (and to some extent the 350~$\mu$m) images show detections of the non-thermal continuum emission previously reported at 870~$\mu$m by \citet{2008A&A...490...77W}.

We model the dust spectral energy distribution using a single component modified blackbody, and find temperatures of $\le 30$~K in the centre that decrease with radius.  Using the temperature map, we find the total dust mass to be $(1.59 \pm 0.05) \times 10^{7}$~M$_{\odot}$.  The total gas mass is $(2.7 \pm 0.2)\times 10^{9}$~M$_{\odot}$, and combining the dust and gas masses we have produced a gas-to-dust ratio map.  The average gas-to-dust ratio is approximately $103 \pm 8$, similar to that of the Milky Way, with an interesting peak to about 275 towards the centre of the galaxy.  After exploring several possible scenarios to explain the high gas-to-dust mass ratio, the most appealing one is that of a correspondance between the high gas-to-dust mass ratio and the ring about 1~kpc in size, which is best fit by a warped tilted ring model by \citet{2006ApJ...645.1092Q} that consists of a deficit of dusty material inwards of the ring.  The deficit of dust may be due to X-ray emission from the central AGN destroying the dust grains or the jets are removing dust from the centre of the galaxy.

\section*{Acknowledgments}
The research of C.~D.~W. is supported by the Natural Sciences and Engineering Research Council of Canada and the Canadian Space Agency.  PACS has been developed by a consortium of institutes led by MPE (Germany) and including UVIE (Austria); KU Leuven, CSL, IMEC (Belgium); CEA, LAM (France); MPIA (Germany); INAF-IFSI/OAA/OAP/OAT, LENS, SISSA (Italy); IAC (Spain). This development has been supported by the funding agencies BMVIT (Austria), ESA-PRODEX (Belgium), CEA/CNES (France), DLR (Germany), ASI/INAF (Italy), and CICYT/MCYT (Spain).  SPIRE has been developed by a consortium of institutes led by Cardiff University (UK) and including Univ. Lethbridge (Canada); NAOC (China); CEA, LAM (France); IFSI, Univ. Padua (Italy); IAC (Spain); Stockholm Observatory (Sweden); Imperial College London, RAL, UCL-MSSL, UKATC, Univ. Sussex (UK); and Caltech, JPL, NHSC, Univ. Colorado (USA). This development has been supported by national funding agencies: CSA (Canada); NAOC (China); CEA, CNES, CNRS (France); ASI (Italy); MCINN (Spain); SNSB (Sweden); STFC and UKSA (UK); and NASA (USA).  HIPE is a joint development by the Herschel Science Ground Segment Consortium, consisting of ESA, the NASA Herschel Science Center, and the HIFI, PACS and SPIRE consortia.  The James Clerk Maxwell Telescope is operated by The Joint Astronomy Centre on behalf of the Science and Technology Facilities Council of the United Kingdom, the Netherlands Organisation for Scientific Research, and the National Research Council of Canada. This research has made use of the NASA/IPAC Extragalactic Database (NED) which is operated by the Jet Propulsion Laboratory, California Institute of Technology, under contract with the National Aeronautics and Space Administration.  This research made use of APLpy, an open-source plotting package for Python hosted at http://aplpy.github.com.  T.~J.~P. thanks the anonymous referee for providing constructive feedback about this work.


\label{lastpage}

\end{document}